# PIV/BOS Synthetic Image Generation in Variable Density Environments for Error Analysis and Experiment Design

Lalit K. Rajendran[1], Sally P. M. Bane[1] and Pavlos P. Vlachos[2]
1: School of Aeronautics and Astronautics, Purdue University, USA
2: School of Mechanical Engineering, Purdue University, USA

## 1 Abstract

We present an image generation methodology based on ray tracing that can be used to render realistic images of Particle Image Velocimetry (PIV) and Background Oriented Schlieren (BOS) experiments in the presence of density/refractive index gradients. This methodology enables the simulation of aero-thermodynamics experiments for experiment design, error, and uncertainty analysis. Images are generated by emanating light rays from the particles or dot pattern, and propagating them through the density gradient field and the optical elements, up to the camera sensor. The rendered images are realistic, and can replicate the features of a given experimental setup, like optical aberrations and perspective effects, which can be deliberately introduced for error analysis. We demonstrate this methodology by simulating a BOS experiment with a known density field obtained from direct numerical simulations (DNS) of homogeneous buoyancy driven turbulence, and comparing the light ray displacements from ray tracing to results from BOS theory. The light ray displacements show good agreement with the reference data. This methodology provides a framework for further development of simulation tools for use in experiment design and development of image analysis tools for PIV and BOS applications. An implementation of the proposed methodology in a Python-CUDA program is made available as an open source software for researchers.

## 2 Introduction

Particle Image Velocimetry (PIV)[1] and Background Oriented Schlieren (BOS)[2] are widely used techniques to investigate complex flows. In PIV, the flow of interest is seeded with particles and the flow velocity is measured by estimating the particle displacements between two successive frames. In BOS, the density gradients in a flow are measured by the apparent shift of a dot pattern viewed through a variable density medium, where the displacement is evaluated using methods similar to PIV. To assess and improve the accuracy of the displacement estimation algorithms, synthetic particle and/or BOS images are required. For the images to be suitable for testing the algorithms, they must be realistic, i.e., they should display real world artifacts like optical aberrations due to the camera setup, out-of-focus effects, etc. To simulate these effects, current synthetic image generation techniques use empirical models which are too generic to be applied to specific optical systems [1]. In addition, these models cannot be used to simulate effects like ray deflection due to the presence of density gradients, which is an important concern in compressible flow experiments.

Ray tracing is a physically realistic alternative, where light rays generated from the particles/dot patterns are traced through the flow under investigation and the optical setup, all the way to the camera sensor. This approach does not require any ad-hoc models and can also naturally handle effects like ray deflection due to density gradients. Although ray tracing tools are ubiquitous across many applications, the methodology presented herein is novel as it is the first to combine density gradients effects, specific user-defined optics, and camera/sensor parameters along with fluid flow in one package tailored for simulating general aero-thermodynamics experiments.

A significant challenge is that ray tracing is computationally expensive due to the large number of rays required to faithfully reproduce an image. For example, a typical tomographic (Tomo) PIV experiment would have about 100,000 particles inside a laser sheet volume of 300 $cm^3$. To simulate a particle with sufficient dynamic range, about 10,000 rays are required, which corresponds to a total of 1 billion rays to render a single image, thus posing a significant computational challenge. However, since the path of each light ray is independent of all other rays, this process can be very efficiently parallelized and implemented on Graphics Processing Units (GPUs) which can launch several thousand threads at a time in addition to about a trillion floating point operations (FLOPs) per second. This capability of GPUs is exploited in the current work to significantly accelerate the image generation process.

In the subsequent sections, we first describe in detail the synthetic image generation methodology used to render realistic particle/BOS images in a varying density/refractive-index medium, and then present an application for Background Oriented Schlieren (BOS) experiments. This approach renders images unique to a given optical setup and can be a valuable tool for guiding the choice of optical elements and their placements in the experimental setup to mitigate adverse effects like optical aberrations and steep viewing angles. On the other hand, these effects can be deliberately included for error analysis so that the robustness of an algorithm can be tested for a wide variety of conditions. Some sample particle images generated using the proposed methodology are shown in Figure 1.

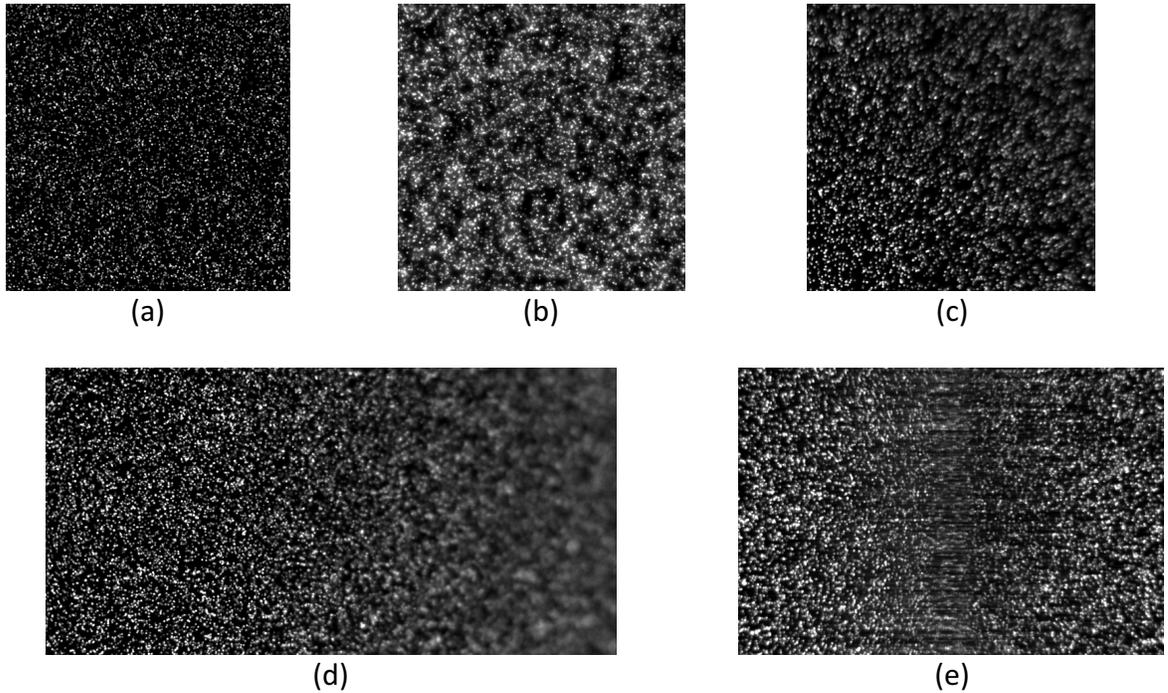

**Figure 1.** Sample particle images generated using the proposed methodology displaying some common experimental artifacts. (a) Normal Image, (b) Out of focus effects, (c) Lens aberration near edges, (d) Perspective effect, (e) Blurring due to density gradients (normal shock wave in the center of the image).

## 3  Image Generation Methodology

The image generation process, shown schematically in Figure 2, is comprised of four steps: (1) generating the light rays, (2) tracing the light rays through density gradients, (3) propagating the light rays through optical elements, and (4) intersecting the rays with the camera sensor to update the pixel intensities. Each of these steps is described in more detail in the following sections.

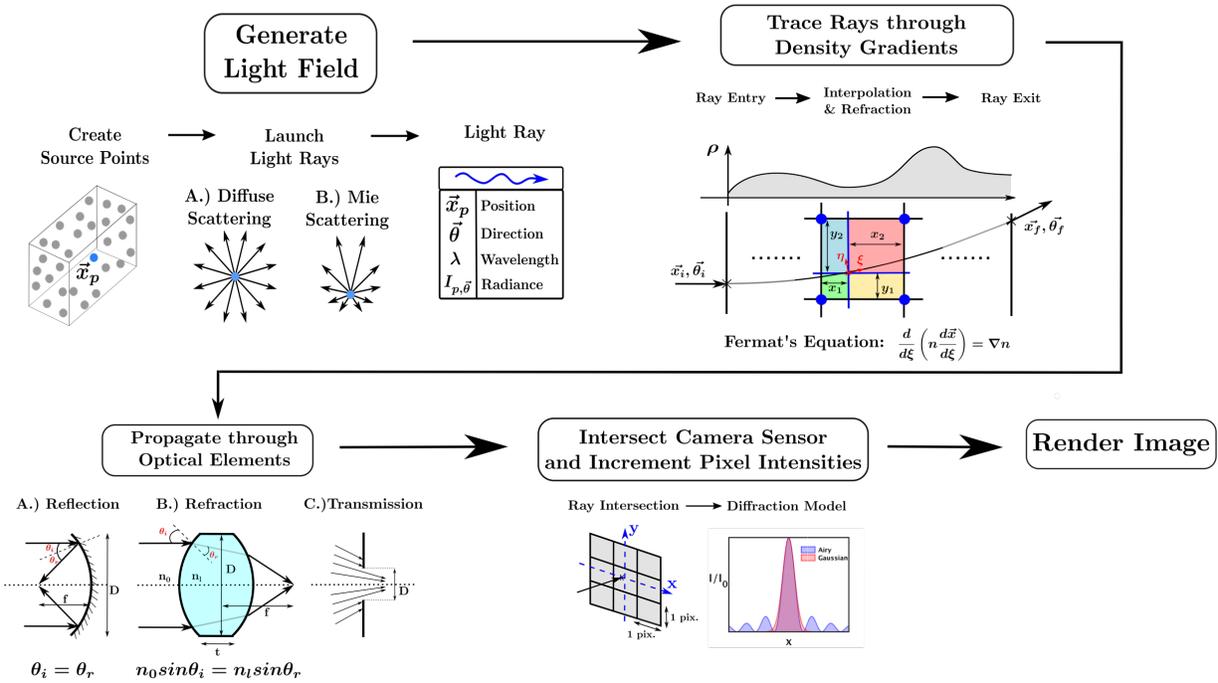

**Figure 2.** Synthetic image generation methodology.

## 3.1 Generating the Light Field

The light rays generated from the particles/dot pattern are considered as vectors connecting source points in the flow field to points of intersection on the camera lens. The source point can be a particle for a PIV experiment or a dot pattern for a BOS experiment. The origin of the light ray vector corresponds to the position of the source point, and its direction corresponds to a unit vector connecting the origin to the point of intersection on the camera lens. Ideally, an infinite number of such light rays can be generated from each source point towards points located on the camera lens. Increasing the number of light rays increases the dynamic range of the generated images but also increases the computational cost.

The radiance of the light ray may have an angular dependence based on the type of scattering associated with the source point. In the case of a PIV particle field where the particle diameters are typically of the same order of the wavelength of the laser, the radiance of the light ray can be estimated using Mie scattering [3]. The scattering cross-section and efficiency depends on the size of the particle, the wavelength of the laser beam, the relative refractive index of the particle with respect to the medium, and the angle between the light ray vector and the direction of propagation of the laser beam. The Mie scattering computations are performed using the method outlined in Bohren & Huffman [4].

## 3.2 Tracing Rays through Density Gradients

A light ray will experience changes in its direction as it passes through a medium containing density gradients due to the dependence of the refractive index on the local density as expressed by the Gladstone-Dale relation:

$$n = K\rho + 1 \qquad (1)$$

where $n$ is the refractive index of the medium, $\rho$ is the density, and $K$ is the Gladstone-Dale constant, which has a value of 0.226 cm$^3$/g for air. Therefore, regions of density gradients also contain refractive index gradients. For a medium containing a continuous change of refractive index, Fermat's principle from geometric optics enables a fast and accurate computation of the trajectory of a light ray through the medium, and the equation for the ray curve is given by [5],

$$\frac{d}{d\xi}\left(n\frac{d\vec{x}}{d\xi}\right) = \nabla n \qquad (2)$$

Here $\vec{x}(\xi)$ represents the ray curve and $(\xi, \eta)$ are the ray-fitted co-ordinates as shown in Figure **2**. Equation (2) is transformed and discretized using a 4$^{th}$ order Runge-Kutta algorithm following the method of Sharma et. al. [6] and the position and direction of the light ray passing through the variable density medium can be updated based on the local refractive index gradient as follows,

$$R_{i+1} = R_i + \left[T_i + \frac{1}{6}(A + 2B)\right]\Delta\xi$$
$$T_{i+1} = T_i + \frac{1}{6}(A + 4B + C) \qquad (3)$$

where $R, T$ are 1D arrays representing the position and direction, respectively, and are given by,

$$R = \begin{pmatrix} x \\ y \\ z \end{pmatrix}; \quad T = n\begin{pmatrix} dx/d\xi \\ dy/d\xi \\ dz/d\xi \end{pmatrix} \qquad . \qquad (4)$$

The variable $n$ is the refractive index and the subscript $i$ represents the grid point corresponding to the given location of the ray. The constants $A$, $B$ and $C$ are functions of the refractive index gradients and are given by,

$$A = D(R_i)\Delta\xi$$
$$B = D\left(R_i + \left(\frac{1}{2}T_n + \frac{1}{8}A\right)\Delta\xi\right)\Delta\xi \qquad (5)$$
$$C = D\left(R_i + \left(T_n + \frac{1}{2}B\right)\Delta\xi\right)\Delta\xi$$

and the function $D$ is given by

$$D = n \begin{pmatrix} \partial n/\partial x \\ \partial n/\partial y \\ \partial n/\partial z \end{pmatrix} = \frac{1}{2} \begin{pmatrix} \partial n^2/\partial x \\ \partial n^2/\partial y \\ \partial n^2/\partial z \end{pmatrix} . \tag{6}$$

An open-source implementation of solving Fermat's equation on a GPU with a piecewise linear approximation (1st order) was provided by SchlierenRay, an artificial schlieren image rendering software developed by Brownlee et. al. [7] Their methodology has been extended to include higher order approximations and integrated with a full light field-based ray tracing approach for the present application.

### 3.3 Propagating Light Rays through Optical Elements

When light rays pass through optical elements, they can undergo one or more of the following processes: (1) reflection (mirrors), (2) refraction (lenses, windows), and (3) selective transmission (apertures). All of theses processes are modeled in the ray tracing methodology, as shown in Figure 2. In all cases, the intersection of a ray with the optical element is first computed based on the element's geometry. For example, in the case of a spherical mirror/lens the intersection point is calculated based on the element center, diameter, and radius of curvature. After computing the intersection, the effect of the element is modeled as follows:

1) Reflection due to mirrors is modeled using the law of reflection based on the direction of the light ray with respect to the local surface normal.

2) Refraction due to lenses/windows is modeled using Snell's Law [8], given by

$$n_i \sin(\theta_i) = n_f \sin(\theta_f), \tag{7}$$

   where $\theta_i$ is the angle of incidence, $\theta_f$ is the angle of refraction, and $n_i$ and $n_f$ are the refractive indices of the two media on either side of the refractive surface. For elements with multiple refractive surfaces like a lens, the refraction is performed sequentially on each surface, considering the possibility of total internal reflection if the ray passes from a medium of higher refractive index to a medium of lower refractive index. It should be noted that this approach is quite general and does not require assumptions regarding the paraxial nature of the light rays (as used in matrix methods) or the thickness of the lens, and it is straightforward to include transmittance and dispersive effects of the lens as required. Further, an array of lenses as in a Plenoptic camera, for example, can also be modeled using this approach.

3) Selective transmission due to apertures is enforced by only allowing light rays that intersect the plane of the aperture and lie within its opening area (or pitch) and blocking the rest.

## 3.4 Intersecting a Ray with the Camera Sensor and Incrementing Pixel Intensities

The final step in the ray tracing process is the intersection of a light ray with the camera sensor, which is solved as a line-plane intersection problem. The diffraction spot is described by an Airy function and is approximated by a Gaussian in this application [9], and the integrated intensity across a pixel is calculated using an error function, as in the case of synthetic PIV image generation [10]. The point of peak intensity is the point of intersection of the light ray with the camera sensor, and the diffraction diameter is a function of the optical system as given by,

$$d_\tau = 2.44\pi f_\#(M + 1)\lambda \qquad . \qquad (8)$$

Here $d_\tau$ is the diffraction diameter, $f_\#$ is the f-number of the camera, $M$ is the magnification, and $\lambda$ is the wavelength of light [9]. For white light illumination in the case of BOS/calibration targets, an effective wavelength corresponding to the green color is used.

This procedure is repeated for all light rays that intersect the camera sensor to obtain an image of the particle field/dot pattern. The dynamic range of the final intensity distribution increases with the number of light rays used to render a particle or dot, but this also increases the computational cost and run time. It was observed from trials that about 10,000 rays are sufficient to provide a 16-bit dynamic range.

## 3.5 Parallelization using CUDA

The ray tracing methodology just described is computationally intensive due to the large number of light rays (~ 1 billion) required to render an image with sufficient dynamic range. Since the trajectories of the light rays are independent of each other, the ray tracing calculations can be parallelized using Graphics Processing Units (GPUs). This methodology was implemented using a CUDA framework with a Python front-end. The images in the present work were generated using an NVIDIA Tesla C2050 GPU, which has 14 streaming multi processors each containing 32 cores for an overall total of 448 cores. Each multi-processor can launch a maximum of 1536 threads amounting to a total of about 21,000 threads at a time.

The details of the parallelization in terms of grids, blocks and threads are as follows. Each thread on the GPU corresponds to a single light ray, and all the computations starting from the ray generation to the intersection with the camera sensor are done independently. All light rays originating from the same particle/dots are organized in blocks, to take advantage of the shared memory in CUDA which has very fast read and write speeds [11]. Thus the information common to all light rays originating from the same particle are stored in shared memory, which frees up the local memory and enables launching a larger number of threads. The number of threads that can be stored in a block and the number of blocks that can be launched are subject to hardware limitations.

## 4 Results

To investigate the accuracy of the overall image generation methodology, we simulate a full Background Oriented Schlieren (BOS) experiment with a known density field and user defined camera optics. We compare the final light ray deflections recorded on the camera sensor to predicted displacements from BOS theory. The density field is taken from a direct numerical simulation (DNS) of homogeneous buoyancy-driven turbulence performed by Livescu et. al. [12]–[14] and was downloaded from the Johns Hopkins University Turbulence Database (JHU-TDB) [12], [15], [16].

Two dimensional (x, y) slices of the flow field from two time instants were chosen, and for each time instant, a three-dimensional density volume was constructed by stacking the same two-dimensional slice along the z-direction, thereby ensuring that the gradient of density in the z direction was zero. This was done to account for the depth integration limitation of BOS measurements and to enable a better comparison of the simulated light ray deflections to theory. The refractive index was calculated using Equation (1) using the Gladstone-Dale constant for air, and the non-dimensional density field from the DNS was scaled by a factor of 1.225 kg/m³ to simulate air properties. The layout of the BOS experimental setup modeled in the image generation software is shown in Figure 3, and the parameters describing the placement of the elements are summarized in Table 1.

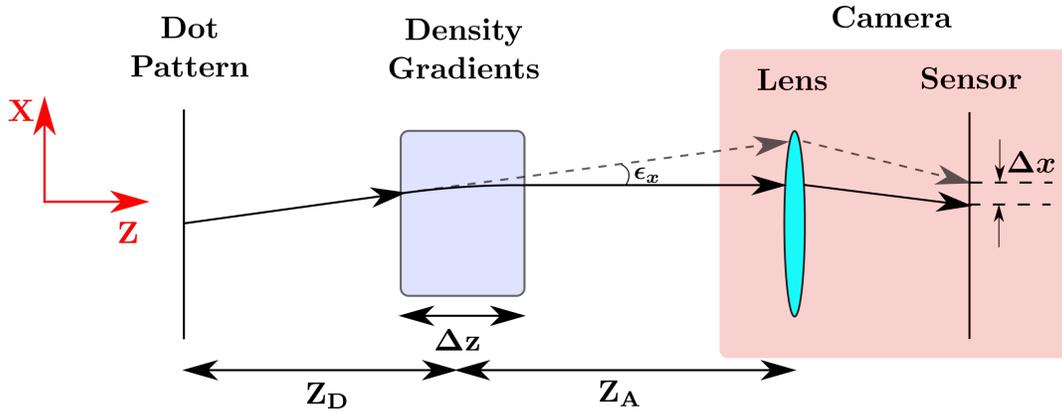

**Figure 3.** Layout of the experimental setup used to simulate a BOS experiment with the DNS density field.

**Table 1.** Summary of image generation parameters used to simulate the BOS experiment shown in Figure 3

| | |
|---|---|
| $Z_A$ | 0.75 m |
| $Z_D$ | 0.25 m |
| $L_x$ x $L_y$ x $L_z$ | 32 x 32 x 10 mm |
| Focal Length | 105 mm |
| Aperture ($f_\#$) | 11 |
| Magnification | 0.12 |
| Pixel Pitch | 10 um |

| Dot density | 20 dots / 32x32 pix. |

The contours of the input density and density gradients, the theoretical displacements, and the light ray displacements from ray tracing simulations are shown in Figure 4. The theoretical displacements for a BOS experiment are given by

$$\Delta \vec{X} = \frac{MZ_D}{n_0} \int_{z_i}^{z_f} \nabla n \, dz$$
$$\approx \frac{MZ_D K}{n_0} (\nabla \rho)_{avg} L_z \qquad (9)$$

where $\Delta \vec{X}$ is the theoretical deflection of a light ray, $(\nabla \rho)_{avg}$ is the path-averaged value of the density gradient, $K$ is the Gladstone-Dale constant, $n_0$ is the ambient refractive index, and $L_z$ is the depth/thickness of the density gradient field [2]. The values of the experimental parameters were taken from Table 1, and the depth averaged density gradient $(\nabla \rho)_{avg}$ is taken to be the two-dimensional density gradient field shown in Figure 4, as identical 2D slices were stacked to create a 3D density field during the simulations.

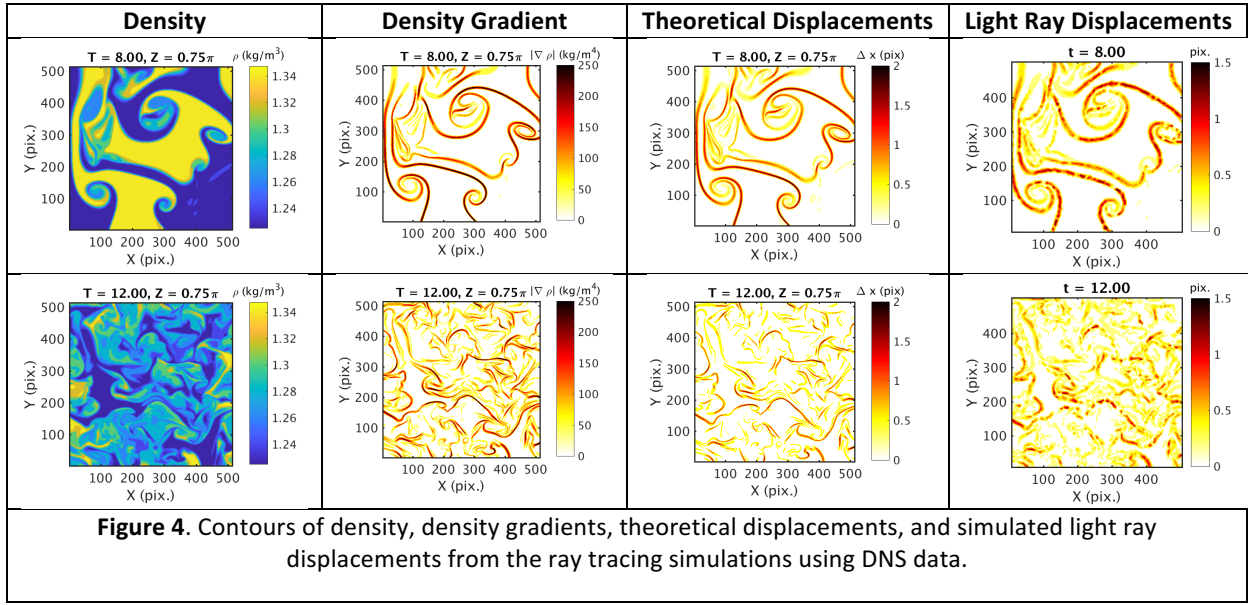

**Figure 4**. Contours of density, density gradients, theoretical displacements, and simulated light ray displacements from the ray tracing simulations using DNS data.

The light ray displacements from the ray tracing simulations will be randomly scattered on the camera sensor due to the random positions of the dots on the target from which the light rays originate. The ray displacements corresponding to a single dot are averaged and interpolated onto a regular grid using a bilinear interpolation and displayed in Figure 4. The figure shows that the contours of light ray displacements from the simulations closely correspond to the theoretical displacements except that they are smoothed out. The mismatch between the theoretical and simulated light ray deflections is due to two reasons: (1) the theoretical equation is based on small angle approximations, and (2) the spatial resolution limitation of the BOS experimental setup whereby the light ray deflection of a dot is the average light ray displacement of all rays comprising a ray cone. Both these effects are consistent with well-known characteristics of BOS experiments [2], [17], [18].

These results, in addition to the sample particle images shown in Figure 1, illustrate the capability of the proposed image generation methodology to accurately generate realistic PIV/BOS images. The methodology thus enables the introduction of experimental artifacts such as optical aberrations and distortions due to density gradient fields into the image generation process in a deliberate and controlled manner.

## 5 Conclusion

An image generation methodology was proposed and implemented to render realistic PIV and BOS images in variable density environments with a user-defined optical setup. The methodology involves generation of light rays from a particle or dot pattern, propagation of the light rays through density gradients using Fermat's equation and a $4^{th}$ order Runge-Kutta scheme, reflection/refraction/transmission of the light rays by optical elements, and intersection of the rays with the camera sensor to update pixel intensities using a diffraction model. The computationally intensive ray tracing process was parallelized and implemented on GPUs using CUDA, resulting in a significant acceleration of the computations. The developed methodology was used to simulate a BOS experiment with a known density field obtained from DNS of buoyancy-driven turbulence. The light ray deflections from the ray tracing show good agreement with the theoretical estimates. This methodology provides a strong framework for further development of simulation tools for use in experiment design by incorporating additional features specific to a given experiment. The methodology can also be a valuable tool for error analysis to study the effect of various elements of an optical setup on the final error, and provide directions to improve image analysis tools for PIV and BOS applications.

## 6 Acknowledgment

This material is based upon work supported by the U.S. Department of Energy, Office of Science, Office of Fusion Energy Sciences under Award Number DE-SC0018156.